\documentclass[11pt]{article}
\usepackage{graphicx}
\usepackage{pstricks}
\newcommand{\be}{\begin{equation}}
\newcommand{\ee}{\end{equation}}
\def\ba{\begin{eqnarray}}
\def\ea{\end{eqnarray}}

\def\({\left(}
\def\){\right)}
\def\ls{\left[}
\def\rs{\right]}

\title{{\bf Noncommutative nonsingular black holes}}
\author{Piero Nicolini\\ {\footnotesize Department of Theoretical Physics,
Jo\v{s}ef Stefan Institute, Ljubljana, Slovenia}\\
{\footnotesize Department  of Mathematics, Polytechnic of Turin,
Turin, Italy}\\ {\footnotesize Department  of Mathematics and
Informatics, University of Trieste, Trieste, Italy
}\\{\footnotesize INFN, National Institute for Nuclear Physics,
Trieste, Italy}
\\
{\footnotesize e-mail: Piero.Nicolini@cmfd.univ.trieste.it}}
\date{\today}

\begin{document}
\maketitle

\begin{abstract}
Adopting noncommutative spacetime coordinates, we determined a new
solution of Einstein equations for a static, spherically symmetric
matter source.  The limitations of the conventional Schwarzschild
solution, due to curvature singularities, are overcome. As a
result, the line element is endowed of a regular DeSitter core at
the origin and of two horizons even in the considered case of
electrically neutral, nonrotating matter. Regarding the Hawking
evaporation process, the intriguing new feature is that the black
hole is allowed to reach only a finite \textit{maximum}
temperature, before cooling down to an \textit{absolute zero}
extremal state. As a consequence the quantum back reaction is
negligible.
\end{abstract}

\section{Introduction}
In spite of  decades of efforts, a complete and satisfactory
quantum theory of Gravity does not yet exist. Thus a great
interest has arisen   towards the class of model theories able to
reproduce quantum gravitational effects, at least in some limit.
Quantum Field Theory on a noncommutative manifold or shortly
Noncommutative Field Theory (NFT) belongs to such class of model
theories. Indeed we retained NFT the low energy limit of String
Theory, which is the most promising candidate to be the quantum
theory of Gravity.

The starting point of the NFT is the adoption of a noncommutative
geometry, namely a manifold whose coordinates may fail to commute
in analogy to the conventional noncommutativity among conjugate
variables in quantum mechanics \be \ls
\textbf{x}^i,\textbf{x}^j\rs = i\
\theta^{ij}\,\,\,\,\,\,\,i,j=1,...,n \label{ncrules} \ee with
$\theta^{ij}$ an antisymmetric (constant) tensor of dimension $($
length $)^2$. Eq. (\ref{ncrules}) provides an uncertainty in any
measurement of the position of a point on the noncommutative
manifold.  Indeed we cannot speak of point anymore but rather of
delocalized position according to the noncommutative uncertainty.
The physical motivation for assuming a noncommutative geometry
relies on the bad short distance behavior of field theories,
gravitation included. In fact this is a typical feature of
theories dealing with point like objects, a problem that has not
been completely solved by String Theory too. NFT could provide the
solution, since a noncommutative manifold is endowed of a natural
cut off due to the position uncertainty. This aspect is in
agreement with the long held belief that spacetime must change its
nature at distances comparable to the Planck scale. Quantum
Gravity has an uncertainty principle which prevents one from
measuring positions to better accuracies than the Planck length,
indeed the shortest physically meaningful length. In spite of this
promising programme and the issue of a seminal paper dated in the
early times \cite{sny}, the interest towards NFT is rather recent.
Indeed a significant push forward was given only when, in the
context of string theory, it has been shown that target spacetime
coordinates become \textit{noncommuting} operators on a $D$-brane
\cite{sw}. This feature promoted the interpretation of NFT, among
the class of nonlocal field theories \cite{Pais:1950za}, as the
low energy limit of the theory of open strings.

The inclusion of noncommutativity in field theory in flat space is
the subject of a large literature. On the contrary, the purpose of
the paper is to introduce  noncommutativity effects in the
gravitational field, with the hope that noncommutativity could
solve the long dated problems of curvature singularity in General
Relativity.  This investigation is motivated by some mysterious
feature of the physics of quantum black holes. Indeed the interest
towards their complete understanding have been increasing since
the remarkable Hawking discovery about the possibility for them to
emit radiation \cite{hawking}. The general formalism employed is
known as quantum field theory in curved space \cite{bd}. Such a
formalism provides, in terms of a quantum stress tensor
$\left<T_{\mu\nu}\right>$ \cite{swave}, a satisfactory description
of black holes evaporation until the graviton density is small
with respect to matter field quanta density. In other words
quantum geometrical effects have to be retained negligible, a
condition that is not more valid in the terminal stage of the
evaporation. The application of noncommutativity to gravity could
provide, in an effective way, the still missing description of
black holes in those extreme regimes, where stringy effects are
 considered relevant \cite{corr,paddy}.

\section{Noncommutative field theory models}

There exist many formulations of NFT, based on different ways of
implementing non local deformations in field theories, starting
from (\ref{ncrules}). The most popular approach is founded on the
replacement of the point-wise multiplication of fields in the
Lagrangian by a non-local Weyl-Wigner-Moyal $\ast$-product
\cite{Moyal}. In spite of its mathematical exactitude, the
$\ast$-product NFT suffers non trivial limitations. The Feynman
rules, obtained directly from the classical action, lead to
unchanged propagators, while the only modifications, concerning
vertex contributions, are responsible of the non-unitarity of the
theory and of UV/IR mixing. In other words UV divergences are not
cured but accompanied by, surprisingly emerging, IR ones. While
unitariety can be restored, the restriction of noncommutative
corrections only to interaction terms is a  non intuitive feature,
which appears in alternative formulations \cite{Bahns:2002vm} too.

Against this background, the coordinate coherent states approach,
based on an oscillator representation of noncommutative spacetime,
leads to a UV finite, unitary and Lorentz invariant field theory
\cite{ae}. The starting point of this formulation is to promote
the relation (\ref{ncrules}) to an equation between Lorentz
tensors \be \ls \textbf{x}^\mu,\textbf{x}^\nu \rs = i\
\theta^{\mu\nu}\hspace{1.5cm}\mu,\ \nu = 0 ,...,n \label{ncx}
 \ee
Thus $\theta^{\mu\nu}$, now an antisymmetric Lorentz tensor, can
be represented in terms of a block diagonal form
\begin{equation}
\theta^{\mu\nu}= diag \left( \hat{\theta }_1 ,\hat{ \theta } _2
,..., \hat{ \theta } _{n/2} \right)\hspace{1.5cm}\mu,\ \nu = 0
,...,n \label{diag}
\end{equation}
where $\hat{ \theta } _i =\theta_i \left(
\begin{array}{cc}
0 & 1 \\
-1 & 0%
\end{array}%
\right) $. In case of odd dimensional manifold the last term on
the diagonal is zero. In other words, for the covariance of
(\ref{ncx}) we provide a foliation of spacetime into
noncommutative planes, defined by (\ref{diag}). There is a
condition to be satisfied: field theory Lorentz invariance and
unitariety implies that noncommutativity does not privilege any of
such planes, namely there is a unique noncommutative parameter
$\theta_1=\theta_2=...\ \theta_{n/2}=\theta$.

The other key ingredient of this approach is the interpretation of
conventional coordinates as mean values of coordinate operators
subject to (\ref{ncrules}), to take into account the quantum
geometrical fluctuations of the spacetime manifold. Mean values
are calculated over coherent states, which results eigenstates of
ladder operator, built with noncommutative coordinates only. In
the absence of common eigenstates, the choice of coherent states
is motivated by the fact that  they result the states of minimal
uncertainty and provide the best resolution of the position over a
noncommutative manifold. In other words, the effective outcome is
the loose of the concept of point in favor of smeared position on
the manifold. At actual level, the delocalization of fields is
realized by deforming the source term of their equations of
motion, namely substituting Dirac delta distribution (local
source) with Gaussian distribution (nonlocal source) of width
$\sqrt{\theta}$. As a result, the ultraviolet behavior of
classical and quantum fields \cite{pnag} is cured.

\section{The noncommutative black hole}

To provide a black hole description by means of a noncommutative
manifold, one should know how to deal with the corresponding
gravity field equations. Fortunately, this nontrivial problem can
be circumvented by a noncommutative deformation of only the matter
source term, leaving unchanged the Einstein tensor. This
procedure, already followed in $(1+1)$ dimensions
\cite{Nicolini:2005de} and $(3+1)$ linearized General Relativity
\cite{Nicolini:2005zi}, is in agreement with the general
prescription to obtain nonlocal field theories from
noncommutativity \cite{ae}. This line of reasoning is supported by
the following motivations. Noncommutativity is an intrinsic
property of a manifold and affects matter and energy distribution,
by smearing point-like objects, also in the absence of curvature.
On the other hand, the metric is a geometrical device defined over
the underlying manifold, while curvature measure the metric
intensity, as a response to the presence of mass and energy
distribution. Being the energy-momentum tensor the tool which
gives the information about the mass and energy distribution, we
conclude that, in General Relativity, noncommutativity can be
taken into account by keeping the standard form of the Einstein
tensor in the l.h.s. of the field equations and introducing a
modified source term in the r.h.s..

Therefore we assume, as mass density of the noncommutative
delocalized particle, the Gaussian function of minimal width
$\sqrt{\theta}$
\begin{equation}
\rho_\theta\left(\,\vec{x}\,\right)=
\frac{M}{\left(\,4\pi\theta\,\right)^{3/2}}\,
\exp\left(-\vec{x}^{\,2}/4\theta\,\right)
 \label{t00}
 \end{equation}
Thus the particle mass $M$ is \textit{diffused} throughout a
region of linear size $\sqrt{\theta}$ taking into account the
intrinsic uncertainty encoded in the coordinate commutator
(\ref{ncx}). The distribution function
$\rho_\theta\left(\,\vec{x}\,\right)$ is static, spherically
symmetric and exponentially vanishing at distances $r>>
\sqrt\theta$. In this limit $\rho_\theta\left(\,\vec{x}\,\right)$
reproduces point-like sources and leads to the conventional
Schwarzschild solution.
%Even if phenomenological results, so far,
%imply that noncommutativity is not visible at presently accessible
%energies, i.e. $\sqrt\theta< 10^{-16}\, cm $, new physics is
%expected at distance $r\simeq \sqrt\theta$  where a non negligible
%density of energy and momentum is present.
On these grounds, we are looking for a noncommutative version of a
 with $T^0_{\,\, 0} = \rho_\theta\left(\,\vec{x}\,\right )$ as
source of Einstein equations. There are two further conditions to
be taken: the covariant conservation of the energy-momentum tensor
$\nabla_\nu\, T^{\mu\nu}=0$ and $g_{00}=-g_{rr}^{-1}$ to preserve
a Schwarzschild-like property. Therefore the solution of Einstein
equations is\footnote[1]{We use convenient units $G_N=1$, $c=1$.}:
\begin{equation}
 ds^2 =\left(\, 1- \frac{4M}{r\sqrt{\pi}}\, \gamma \right)\, dt^2 - \left(\,
1-\frac{4M}{r\sqrt{\pi}}\, \gamma \right)^{-1}\, dr^2 -
r^2\,d\Omega^2\label{ncs}
\end{equation}
where $\gamma\equiv\gamma\left(3/2 \ , r^2/4\theta\,
\right)=\int_0^{r^2/4\theta} dt\, t^{1/2} e^{-t} $ is the lower
incomplete Gamma function. This line element describes the
geometry of a noncommutative black hole  and should give us useful
insights about  possible noncommutative effects on Hawking
radiation.
%Rather than a massive, structureless point, a source turns out to
%a \textit{self-gravitating}, \textit{droplet } of
%\textit{anisotropic fluid} of density $\rho_\theta$, radial
%pressure $p_r= -\rho_\theta$ and \textit{tangential pressure}
%\begin{equation}
%p_\perp = -\rho_\theta
%-r\,\partial_r\rho_\theta\left(\,r\,\right)\label{perp}
%\end{equation}
%On physical grounds, a non-vanishing radial pressure is needed to
%balance the inward gravitational pull, preventing droplet to
%collapse into a matter point. This is the basic physical effect on
%matter caused by spacetime noncommutativity and the origin of all
%new physics at distance scale of order  $\sqrt\theta$.
\begin{figure}[h]
\begin{center}
\includegraphics[totalheight=6cm,angle=0]{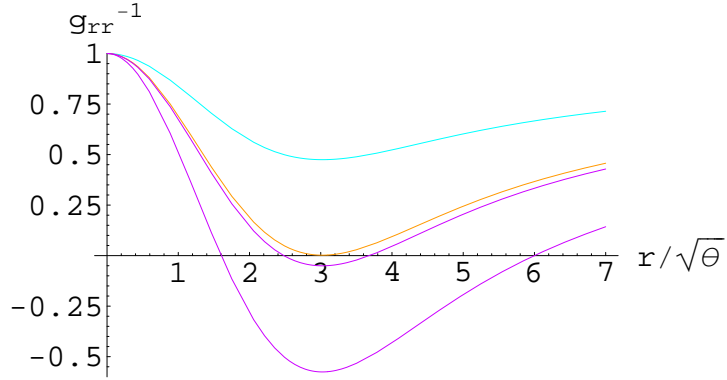}
\caption{\label{hor} $g_{rr}^{-1}$ vs $r$, for various values of
$M/\sqrt\theta$. Intercepts on the horizontal axis give radii of
the event horizons. $M= \sqrt\theta $, ( cyan curve ) no horizon;
$M=1.9\, \sqrt\theta $, (yellow curve) one \textit{degenerate}
horizon $r_0\approx 3.0\, \sqrt\theta$, \textit{extremal}
 black hole;  $M= 3\, \sqrt\theta $ (magenta curve) two horizons.}
\end{center}
\end{figure}
Let's start our analysis from the presence of eventual event
horizons. Since in our case, the equation $g_{00}\left(\,
r_H\,\right)=0$ cannot be solved in closed form, one can
numerically  determine their radius  by plotting $g_{00}$. Figure
(\ref{hor}) shows that noncommutativity introduces new behavior
with respect to standard Schwarzschild black hole. Instead of a
single event horizon, there are different possibilities: (a) two
distinct horizons  for $M> M_0$ (yellow curve); (b) one degenerate
horizon  in  $r_0=3.0\times \sqrt\theta $, with
 $M= M_0=1.9\times (\sqrt\theta) /G $
 corresponding to \textit{\cyan extremal black hole} (cyan curve);
(c) no horizon  for  $M< M_0$ (violet curve).
%Notice that the
%outer horizon has always a radius larger than the extremal limit,
%i.e. $r_H\ge 3\sqrt\theta$, thus our previous approximation
%$r_H\ge 4\sqrt\theta$ is quite acceptable to get a qualitative
%picture of black hole evaporation.
%A more refined treatment is
%only required close to the extremal limit if one need precise quantitative estimates.  \\
In view of these results, there can be no black hole if the
original mass is less than the \textit{minimal mass} $M_0$.
Furthermore,  contrary to the usual case, there can be \textit{two
horizons} for large masses. By increasing $M$, i.e. for $M>> M_0$,
the  \textit{inner  horizon} shrinks to zero, while the
\textit{outer} one approaches the Schwarzschild value $r_H=2M$.
%Equation (\ref{horizon}) can be conveniently rewritten in terms of
%the upper incomplete Gamma function as
%\begin{equation}
% r_H\approx 2M\,\left[\, 1 -\frac{2}{\sqrt{\pi}}\, \Gamma\left(\, 3/2\ ,
%M^2/\theta\, \right) \,\right] \label{horizon2}
%\end{equation}
%The first term in (\ref{horizon2}) is the Schwarzschild radius,
%while the second term brings in $\theta$-corrections. \\
%Two regions are of interest.
%In the ``large radius'' regime $r^2_H/4\theta>>1$ equation
%(\ref{horizon2}) can be solved by iteration. At the first order in
%$M/\sqrt\theta$, we find
%\begin{equation}
%r_H \simeq 2M\,\left(\, 1 - \frac{M}{\sqrt{\pi\theta}}\,
%e^{-M^2/\theta}\,\right)\label{2m}
%\end{equation}
%The effect of noncommutativity is exponentially small, which is
%reasonable to expect since at large distances spacetime can be
%considered as a smooth classical manifold.
% On the other hand, at short distance one expects significant changes due to
%the spacetime fuzziness, i.e. the ``quantum geometry''  becomes
%important $r_H\simeq\sqrt\theta  $. \\

For what concerns the covariant conservation of $T^{\mu\nu}$, one
finds that such requirement leads to
\begin{equation}
T^\theta{}_\theta\equiv \partial_r\left(\, r T^r{}_r\,\right)= -
\rho_\theta\left(\,r\,\right)-
r\,\partial_r\rho_\theta\left(\,r\,\right).
\end{equation}
The emerging picture of is that of a \textit{self-gravitating},
\textit{droplet } of \textit{anisotropic fluid} of density
$\rho_\theta$, radial pressure $p_r= -\rho_\theta$ and
\textit{tangential pressure} $ p_\perp = -\rho_\theta
-r\,\partial_r\rho_\theta\left(\,r\,\right)$. We are not dealing
with a massive, structure-less point. Thus results reasonable that
a non-vanishing radial pressure balances the inward gravitational
pull, preventing droplet to collapse into a matter point. This is
the basic physical effect on matter caused by spacetime
noncommutativity and the origin of all new physics at distance
scale of order  $\sqrt\theta$. Regarding the physical
interpretation of the pressure, we underline that it does not
correspond to the inward pressure of outer layers of matter on the
core of a ``star'', but to a totally different quantity of
``quantum'' nature. It is the outward push, which is
conventionally defined to be negative, induced by noncommuting
coordinate quantum fluctuations. In a simplified  picture,  such a
quantum pressure is the relative of the  cosmological constant  in
DeSitter universe. As a consistency check of this interpretation
 we are going to show that line element (\ref{ncs}) is well described near the
 origin by a DeSitter geometry.

\begin{figure}[h]
\begin{center}
 \includegraphics[totalheight=6cm, angle=0 ]{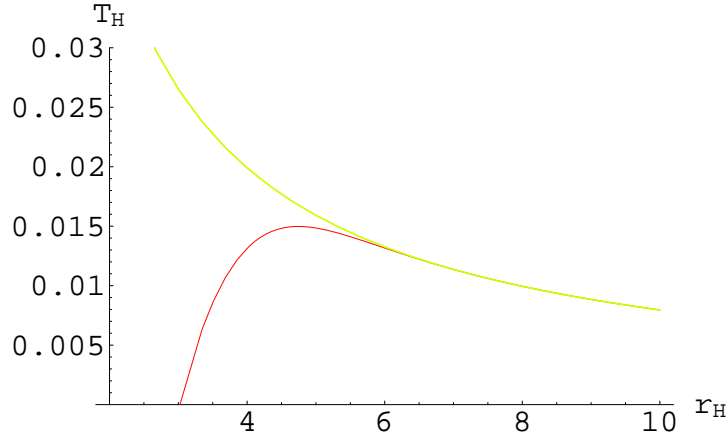}
\end{center}
\caption{\label{T} Plot of $T_H$ vs $r_H$, in $\sqrt\theta$ units.
yellow curve is the plot of (\ref{thnc}): $T_H=0$ for
$r_H=r_0=3.0\sqrt{\theta}$,i.e. for the extremal black hole,
 while the maximum temperature $T_H\simeq 0.015\times
1/\sqrt{\theta}$ corresponds to a mass $M\simeq 2.4
\times\sqrt{\theta}$. For comparison, we plotted in yellow the
standard Hawking temperature.  The two temperatures coincide for
$r_H > 6\,\sqrt\theta$.}
\end{figure}

Let us now consider the black hole temperature $T_H\equiv
\left(\,\frac{1}{4\pi}\, \frac{d g_{00}}{dr}\right)_{r=r_H} $:
\begin{equation}
T_H = \frac{1}{4\pi\,r_H}\left[\, 1
 -\frac{r^3_H}{4\,\theta^{3/2}}\,
  \frac{e^{-r^2_H/4\theta}}{\gamma\left(\, 3/2\ ; r^2_H/4\theta \right)}
\,\right] \label{thnc}
\end{equation}
For large black holes, i.e. $r_H^2/4\theta>>1$, one recovers the
standard result for the Hawking temperature $ T_H= \frac{1}{4\pi\,
r_H} \label{th}$. At the initial state of radiation the black hole
temperature increases while the horizon radius is decreasing. It
is crucial to investigate what happens as $r_H\to \sqrt\theta$. In
the standard (~commutative~) case
  $T_H$ diverges and this puts limit on the
validity of the conventional description of Hawking radiation.
Against this scenario, temperature (\ref{thnc}) includes
noncommutative effects which are relevant at distances comparable
to  $\sqrt\theta$. Behavior of the temperature $T_H$ as a function
of the horizon radius is plotted in Fig.(\ref{T}).
In the region $r_H\simeq \sqrt\theta $, $T_H$ deviates from the
standard hyperbola. Instead of exploding with shrinking $r_H$,
$T_H$ reaches a maximum
%\footnote[2]{We are in the range of
%distances, $r_H\ge 4\sqrt\theta $, where our approximation is
%physically reliable.}
in $r_H\simeq 4.7\sqrt\theta $
corresponding to a mass $M\approx 2.4\, \sqrt\theta/ G_N$, then
quickly drops to zero for $r_H=r_0=3.0\sqrt{\theta}$ corresponding
to the radius of the extremal black hole in figure (\ref{hor}). In
the region $r< r_0 $ there is no black hole  and the corresponding
temperature cannot be defined. As a summary of the results, the
emerging picture of non commutative black hole is that for $M >>
M_0$ the temperature is given by the Hawking temperature
(\ref{th}) with negligibly small exponential corrections, and
increases, as the mass is radiated away. $T_H$ reaches a maximum
value at $M= 2.4\, \sqrt\theta$ and then drops  as $M$ approaches
$M_0$. When $M=M_0$, $T_H=0$, event horizon is degenerate, and we
are left with a ``frozen'' extremal black hole.

At this point, important issue of Hawking radiation back-reaction
should be discussed. In commutative case one expects relevant
back-reaction effects during the terminal stage of evaporation
because of huge increase of temperature \cite{swave,backr}. As it
has been shown, the role of noncommutativity is to cool down the
black hole in the final stage. As a consequence, there is a
suppression of quantum back-reaction since the black hole emits
less and less energy. Eventually, back-reaction may be important
during the maximum temperature phase. In order to estimate its
importance in this region, let us look at the thermal energy
$E=T_H\simeq 0.015\, /\sqrt{\theta}$ and the total mass $M\simeq
2.4\,\sqrt{\theta}\,M_{Pl.}^2  $. In order to have significant
back-reaction effect $ T_H^{Max}$ should be of the same order of
magnitude as $M$. This condition leads to the estimate
$\sqrt{\theta}\approx 0.2\, l_{Pl.}\sim 10^{-34}\, cm
\label{stima}$. Expected values of $\sqrt{\theta}$ are well above
the Planck length $l_{Pl.}$, while the back-reaction effects are
suppressed even if $\sqrt{\theta}\approx 10\, l_{Pl.}$ and
$T_H^{Max}\approx 10^{16}\, GeV$.  For this reason we can safely
use unmodified form of the metric (\ref{ncs}) during all the
evaporation process.

 Finally, we would like to clarify what
happens if the starting object has mass smaller than $M_0$, with
particular attention to the eventual presence of  a \textit{naked
singularity}. To this purpose we are going to study the curvature
scalar near $r=0$. The short distance behavior of $R$ is given by
\begin{equation}
 R\left(\, 0\,\right)=\frac{4M}{\sqrt\pi\, \theta^{3/2}}
  \label{ricci0}
\end{equation}
For $r<<\sqrt\theta$ the curvature is actually \textit{constant
and positive}. Thus, an eventual naked singularity is replaced
 by a DeSitter, \textit{regular} geometry around the origin.  Earlier attempts
 to avoid curvature singularity at the origin of
 Schwarzschild metric have been made by matching DeSitter
 and Schwarzschild geometries both along time-like \cite{olds}, and
space-like matter shells \cite{dscore}, or constructing regular
 black hole geometries by-hand \cite{regbh}. In our approach, it is
 noncommutativity that induces a smooth and continuous transition between
 the two geometries.

 \section{Concluding remarks}
The above results show that the coordinate coherent state approach
to noncommutative effects can cure the singularity problems at the
terminal stage of black hole evaporation.

In particular we have shown that noncommutativity, being an
intrinsic property of the manifold itself, can be introduced in
General Relativity by modifying the matter source. The
Energy-momentum required for this description is of form of the
ideal fluid, although a non-trivial pressure is invoked. In spite
of complicated equation of state it can be studied in the regions
of interest and new black hole behavior is discovered in the
region $r\simeq \sqrt\theta$. Specifically, we have shown that
there is a minimal mass $M_0= 1.9\, \sqrt{\theta}$ to which a
black hole can decay through Hawking radiation.  The reason why it
does not end up into a naked singularity is due to the finiteness
of the  curvature at the origin. The everywhere regular  geometry
and the residual  mass $M_0$ are both manifestations of the
Gaussian delocalization of the source in the noncommutative
spacetime. On the thermodynamic side,  the same kind of
regularization  takes place eliminating the divergent behavior of
Hawking temperature. As a consequence there is a maximum
temperature that the black hole can reach before cooling down to
absolute zero. As already anticipated in the introduction,
noncommutativity regularizes divergent quantities in the final
stage of black hole evaporation in the same way it cured UV
infinities in noncommutative quantum field theory. We have also
estimated that back-reaction does not modify the original metric
in a significant manner.

\section*{Acknowledgements}
The author thanks the ``Dipartimento di Fisica Teorica
dell'Universit\`a di Trieste'', the PRIN 2004 programme ``Modelli
della teoria cinetica matematica nello studio dei sistemi
complessi nelle scienze applicate'' and the CNR-NATO programme for
financial support.

\end{document}